%% file: pap_2pi_tac_rev.tex
\documentclass[10pt,a4paper]{JHEP3}
\usepackage[dvips]{graphicx}
\usepackage{epsfig}
\usepackage{amssymb}
\input epsf.tex
\usepackage{amsmath}
\newcommand{\intveck}{\int \frac{d^3 k}{(2\pi)^3}\,}
\newcommand{\intvecklat}{\int_{-\pi/a_s}^{\pi/a_s} \frac{d^3 k}{(2\pi)^3}\,}

\newcommand{\bal}{\begin{align}}
\newcommand{\eal}{\end{align}}
\newcommand{\bi}{\begin{itemize}}
\newcommand{\ei}{\end{itemize}}
\newcommand{\beas}{\begin{eqnarray*}}
\newcommand{\eeas}{\end{eqnarray*}}

\newcommand{\be}{\begin{eqnarray}}
\newcommand{\ee}{\end{eqnarray}}
\newcommand{\beq}{\begin{equation}}
\newcommand{\eeq}{\end{equation}}

\input{newcomm}
\usepackage[dvips]{graphicx}

\title{Tachyonic preheating using 2PI-1/N dynamics and the classical approximation}

\author{Alejandro Arrizabalaga, Jan Smit and Anders Tranberg\\
Institute for Theoretical Physics, University of Amsterdam, \\
       Valckenierstraat 65, 1018 XE Amsterdam, the Netherlands.\\
}


\preprint{ITFA-2004-35}

\abstract{We study the process of tachyonic preheating
using approximative quantum equations of motion
derived from the 2PI effective action.  The 
$O(N)$ 
scalar (Higgs) field is
assumed to experience a fast quench
which is represented by an instantaneous flip of the sign of the
mass parameter. The equations of motion are solved numerically on
the lattice, and the Hartree and 1/N-NLO approximations are
compared to the classical approximation. Classical dynamics is
expected to be valid, since the occupation numbers 
can rise to large values
during tachyonic preheating. We find that the classical
approximation
performs excellently at short and intermediate times, even
for couplings in the larger region currently allowed for the
SM Higgs. This is
reassuring, since all previous numerical studies of tachyonic
preheating and baryogenesis during tachyonic preheating have used
classical dynamics.
We also compare different initializations for the classical
simulations.}

\begin{document}

%
%

\section{Introduction\label{intro}}

In models of hybrid inflation \cite{Shafi:bd,Linde:1993cn}, one scalar field (the inflaton, $\sigma$) triggers the symmetry breaking of a second (the Higgs field, $\phi$)\footnote{Although our main emphasis is on the Standard Model Higgs, the discussion applies to a general scalar field.} at the end of or after inflation \cite{Copeland:2001qw,vanTent:2004rc}. Typically the inflaton-Higgs coupling is introduced as an effective contribution to the quadratic Higgs term,
\begin{align}
\frac{1}{2}\mu^{2}_{\rm eff}\phi^{2}=\frac{1}{2}\left(\tilde{\mu}^{2}-\lambda_{\sigma\phi}\sigma^{2}\right)\phi^{2}.
\end{align}
As the inflaton goes down, this effective mass $\mu^{2}_{\rm eff}(t)$
changes sign from positive to negative, and as a result the Higgs field
performs symmetry breaking by ``rolling'' into its broken phase minimum. The rolling off its unstable maximum
is accompanied by an exponential growth of the amplitude of the momentum modes
of the Higgs with $\mu_{\rm eff}^2(t)+ \veck^2 < 0$,
a process known as ``spinodal instability'' or ``tachyonic preheating'' \cite{Felder:2000hj}.
Interactions end the transition, and the energy
becomes redistributed through scattering into all the modes and other fields coupled to the Higgs. Eventually the system will thermalize to some temperature, which is called the reheating temperature of the Universe after inflation.
Symmetry breaking and preheating are nonperturbative out-of-equilibrium processes, which are difficult to treat analytically 
\cite{Boyanovsky:1993vi,Boyanovsky:1993pf}.

Numerical studies of tachyonic preheating \cite{Felder:2000hj,Felder:2001kt,Rajantie:2000nj,Copeland:2002ku,Borsanyi:2003ib,Skullerud:2003ki} and baryogenesis during tachyonic preheating \cite{Garcia-Bellido:2003wd,Tranberg:2003gi} apply the classical approximation to the out-of-equilibrium field dynamics, in which the evolution is determined by the classical equations of motion. The argument for making this approximation is that the important unstable low momentum modes acquire large occupation numbers; see \cite{Garcia-Bellido:2002aj,Smit:2002yg} for a detailed justification.

Whereas the instability and growth of the scalar field modes is due to the mass squared $\mu_{\rm eff}^{2}$ becoming negative, it is not obvious what happens when the Higgs is coupled to gauge fields. In \cite{Skullerud:2003ki,Garcia-Bellido:2003wd} it was seen that the low-momentum modes of the gauge field also acquire large occupation numbers,
supporting the classical approximation for the coupled gauge-Higgs system.
This large growth is essential for successful baryogenesis during (electroweak) 
tachyonic preheating 
\cite{Garcia-Bellido:1999sv,Krauss:1999ng,Copeland:2001qw,Garcia-Bellido:2003wd,Tranberg:2003gi}, since these low-momentum modes are supposed to dominate the dynamics of the anomalous baryon number violating processes in the Standard Model (SM)(for a review, see \cite{Rubakov:1996vz}).

In an attempt to use a {\em quantum} description, while still incorporating
the inhomogeneous nature of the typical non-perturbative field configurations
of defects and anomalous baryon number violations, fermions
in classical backgrounds were studied in \cite{AaSm99,Aarts:1999zn}, and scalar
fields in the Hartree approximation with inhomogeneous mean fields
\cite{SaSm00a,BePa01,Salle:2002fu,Salle:2003ju}, in 1+1 dimensions.

In this paper we apply an alternative description of out-of-equilibrium 
dynamics of quantum fields, using 2PI effective action techniques 
\cite{Cornwall:1974vz,BeCo01}. In this approach equations of motion for quantum 
correlators are derived from a suitable quantum effective action (the 2PI 
effective action), in
which the basic variables are the one- and two-point functions.
Although the approach is formally exact, practical calculations
require us to truncate an infinite expansion of loop diagrams ($\Phi$), 
defining a truncated 2PI effective action.
The stationarity of the truncated effective action provides the $\Phi$-derived 
equations of motion.
These equations include infinite loop resummations in the same manner as 
truncations of the Schwinger-Dyson hierarchy, and the two are closely related 
\cite{Alexthesis} 
(for other related work see 
\cite{Cooper:1997ii,Blagoev:2001ze,Cooper:2002qd,Cooper:2002ze,Mihaila:2003mh}).

A crucial feature of $\Phi$-derivable approximations is that the derived equations of motion conserve global symmetries and associated Noether currents in time. In particular the energy is conserved. This is a useful feature when studying out-of-equilibrium processes where most other quantities such as particle distribution functions evolve in a complicated way. The conservation of energy applies not only to the full effective action, but to any level of truncation.

In \cite{BeCo01} the formalism was numerically implemented for
a relativistic scalar field in 1+1 dimensions for a study of the approach to equilibrium. Subsequent studies include \cite{Aarts:2001qa,Berges:2001fi}
in 1+1 dimensions, \cite{Juchem:2003bi} in 2+1 and \cite{Berges:2002wr,Berges:2004ce} for fermions and scalars in 3+1 dimensions. The quantum $\Phi$-derivable scheme was compared to full and $\Phi$-derived classical approximations in \cite{Aarts:2001yn}. Finally, the process of resonant preheating was studied in \cite{Berges:2002cz}. 

We report here on a study of
tachyonic preheating in an $O(4)$ model of scalar fields similar to the
Higgs sector of the SM. However, since there are no gauge fields present, we expect Goldstone modes to appear.
In the SM these would turn into the longitudinal modes of the massive gauge bosons. The results presented here may also have a bearing on
effective pion dynamics in heavy-ion collisions.

In particular we will be interested in comparing results from the $\Phi$-derived equations to those obtained in the classical approximation. Initially we tried truncations of $\Phi$ based on the weak
coupling expansion, but the results were problematic because we
run into instabilities. We record here results based on an
expansion in $1/N$, with $N$ the number of fields (here $N=4$),
following \cite{Berges:2001fi,Aarts:2002dj}. This expansion is
systematic and the numerical algorithm is stable even for very
large occupation numbers \cite{Berges:2001fi,Berges:2002cz}.

%
%

\section{Tachyonic preheating\label{sectacpre}}

The system is described by the classical action
\begin{align}
\label{contact}
S=-\int d^{4}x\bigg[ \frac{\partial_{\mu}\phi_{a}(x)\partial^{\mu}\phi_{a}(x)}{2}-\frac{\mu^{2}}{2}\phi_{a}(x)\phi_{a}(x)+\frac{\lambda}{24 N}\left(\phi_{a}(x)\phi_{a}(x)\right)^{2} + V_0 \bigg],
\end{align}
with $V_0 = 3N\mu^4/2\lm$ the energy density at $\ph=0$, and
$a=1,\cdots,N$, summation implied. In equilibrium at zero temperature, the system is in a broken-symmetry phase if $\mu^{2}>0$, with
\be
\bar{\phi}_{a}\equiv \langle \ph_a\rangle =v\, n_a,
\quad
v=\sqrt{\frac{6\,N\,\mu^2}{\lambda}},
\label{vest}
\ee
where $n_a$ is a unit vector, $n^2=1$. The particles in this phase are
$N-1$ massless Goldstone bosons corresponding to the modes
transverse to $n_a$, and the Higgs boson
corresponding to the longitudinal mode, with mass $m = \sqrt{2}\, \mu$.
A feeling for the strength of the interaction
may be obtained from the amplitude for the scattering of Goldstone
bosons $a+b \to c+ d$ ($a,\ldots, d = 1, \ldots, N-1$, assuming
$n_a=\dl_{a,N}$):
\be
{\cal M} = \frac{\lm}{3N}\, \dl_{ab}\dl_{cd} \,
\frac{(p_a+p_b)^2}{(p_a+p_b)^2 +m^2} + \mbox{2 perm.}
= \dl_{ab}\dl_{cd}\left[\frac{(p_a+p_b)^2}{v^2} + O(p^4)\right]
+ \mbox{2 perm.}.
\label{GBscat}
\ee
This leads to an S-wave unitarity bound $\lm < 48\pi N/(N+1)$
\cite{Luscher:1988gc}. For $N=4$, the triviality bound obtained with the lattice
regularization is about two-thirds of this, $\lm \lesssim 80$
\cite{Luscher:1988gc,Gockeler:1992zj}.
In the Standard Model, $v=246$ GeV, and a Higgs mass $m = 175$ GeV 
would give $\lm = 6$. This looks like a weak coupling,
but such a Higgs mass is near the upper end of the interval set by the
constraints from radiative corrections \cite{pdg}.\footnote{ On the
other hand, a potential application to heavy-ion collisions using
effective pion dynamics, where the role of the Higgs is played by
the sigma-resonance, $m \to m_\sg \approx 600$ MeV,
$v\to f_\pi =93$ MeV, would require very strong coupling, $\lm \approx 120$.}

The coupling of the model to the inflaton may be described by the replacement
$-\mu^2 \to \mu_{\rm eff}^2$ in the action (\ref{contact}).
The spinodal instability occurs when the coefficient of the quadratic term in the potential flips sign. 
Ref.\ \cite{Garcia-Bellido:2002aj} contains a detailed analysis of the initial subsequent development, and a justification is given of a classical approximation after the instability has sufficiently progressed; see also \cite{Baacke:2003bt}.
We go to the limit where the flip happens infinitely fast, a quench, which we model by
\be
\label{thequench}
\mu_{\rm eff}^{2}(t<0)=\mu^2,\qquad \mu_{\rm eff}^{2}(t>0)=-\mu^2.
\ee
We
assume that the initial state of the system is the
ground state
in a quadratic potential just before the quench,
a free ``vacuum'' $|0\rangle_0$,
with
correlators\footnote{We use the Fourier decomposition
$\phi(t,{\bf x})=L^{-3/2}\sum_{\bf k}\phi_{\bf k}(t)e^{i{\bf kx}}$ in a
periodic box of linear size $L$.}
\begin{align}
\label{init1}
\langle 0|\phi^{a}_{\bf k}\phi^{b}_{-\bf k}|0 \rangle_0=&\,
\frac{1}{2\sqrt{\mu^2 + k^2}}\delta_{ab},\\
\langle 0|\pi^{a}_{\bf k}\pi^{b}_{-\bf k}|0 \rangle_0=&\,
\frac{\sqrt{\mu^2 + k^2}}{2}\delta_{ab},\\
\langle 0|\phi^{a}_{\bf k}\pi^{b}_{-\bf k}|0 \rangle_0=&\,\frac{i}{2}\delta_{ab},
\label{init3}
\end{align}
where $k=|\veck|$.
As detailed in \cite{Smit:2002yg}, we can then solve for the evolution of these
correlators in the quadratic approximation ($\lambda=0$) as a function of time.
The result is reproduced here for convenience,
%
\begin{align}
\label{eqC1}
C_k^{\ph\ph}(t)\delta_{ab} =&
\langle 0|\phi^{a}_{\bf k}(t)\phi^{b}_{\bf -k}(t)|0\rangle_0
\equiv 
\frac{n_k(t) + 1/2}{\om_k(t)}\,\dl_{ab}\notag \\
=&\frac{1}{2\omega_{k}^{+}}
\left[1+\left(\frac{\omega_{k}^{+2}}{\omega_{k}^{-2}}-1\right)
\sin^2(\om_k^- t)\right]\delta_{ab},
\\
\label{eqC2}
C_k^{\pi\pi}(t)\delta_{ab}=&
\langle 0|\pi^{a}_{\bf k}(t)\pi^{b}_{\bf -k}(t)|0\rangle_0
\equiv [n_k(t) + 1/2]\,\om_k(t)\, \dl_{ab}\notag\\
=&\frac{\omega_{k}^{- 2}}{2\omega_{k}^{+}} \left[1+
\left(\frac{\omega_{k}^{+2}}{\omega_{k}^{-2}}-1\right)
\cos^2(\om_k^- t)\right]\delta_{ab},
\\
\label{eqC3}
C_k^{\ph\pi}(t)\delta_{ab}=&
\langle 0|\ph^{a}_{\bf k}(t)\pi^{b}_{\bf -k}(t)|0\rangle_0
\equiv [\tilde n_k(t) + i/2]\, \dl_{ab}\notag\\
=&\left[
\frac{\omega_{k}^{-}}{4\omega_{k}^{+}}
\left(\frac{\omega_{k}^{+2}}{\omega_{k}^{-2}}-1\right)
\sin(2\om_k^- t) + \frac{i}{2}\right]\delta_{ab},
\end{align}
with $\omega^{\pm}_{k}=\sqrt{\pm\mu^{2}+k^{2}}$.
The derived occupation numbers and frequencies are
\be
\label{spinnk}
n_{k}(t)+1/2=\sqrt{C_k^{\pi\pi}(t)C_k^{\ph\ph}(t)},\qquad \omega_{k}(t)=\sqrt{C_k^{\pi\pi}(t)/C_k^{\ph\ph}(t)},
\qquad \tilde n_k(t) = \mbox{Re}\, C_k^{\ph\pi}(t).
\ee
It follows that modes with $k^2<\mu^2$ grow exponentially with
\be
n_{k}(t)\propto\exp{(2\sqrt{\mu^2-k^{2}}\,t)}.
\ee
We will use equations (\ref{eqC1}) -- (\ref{eqC3}) for comparison at early times.

Also in more general situations \cite{SaSm00a,SaSm00b,Salle:2002fu,Skullerud:2003ki} the above $\om_k$ give an instantaneous characterization of frequency, assuming a translation and rotation-invariant density matrix, and they provide for a corresponding definition of annihilation operators as (concentrating on one real field)
$a_{\veck} = \frac{1}{\sqrt{2\om_k}}\, (\om_k \ph_\veck + i \pi_\veck)$,
and creation operators $a_\veck^{\dagger}$, such that
$\langle a_\veck^{\dagger} a_\veck \rangle = n_k$.
Hence the $n_k$ have a natural interpretation as occupation numbers.
For free fields in equilibrium the $n_k$ and $\om_k$ coincide with the standard
particle numbers and energies. 
The $\tilde n_k$ are given by $\mbox{Im}\,\langle a_\veck a_{-\veck}\rangle = \tilde n_k$. For large $\tilde n_k$ the non-zero commutator corresponding to the imaginary part of (\ref{eqC3}) becomes unimportant in expectation values of suitable observables (e.g.\ particle numbers), which suggests that the classical approximation should be good.
Ref.\ \cite{Garcia-Bellido:2002aj} stresses that $F_k \equiv \tilde n_k \gg 1$ 
is the important criterion for classical behavior. 
For our initial state $n_k$ and $\tilde n_k$ are not independent, but satisfy $(n_k + 1/2 + \tilde n_k)(n_k + 1/2 - \tilde n_k) = 1/4$, so with $\tilde n_k = \sqrt{n_k^2 + n_k}$, $\tilde n_k > \sqrt{2}$ whenever $n_k > 1$, and for large $n_k$ the two become very close. For precise formulations of the classical approximation in terms of probability distributions of initial conditions, see \cite{Garcia-Bellido:2002aj,Smit:2002yg}.

The initial state is $O(N)$ symmetric, which implies that
the mean field stays equal to zero also for the general case $\lm \neq 0$,
\be
\bar{\phi}_{a}(x)=0, \qquad a=1,\cdots, N,
\ee
and that the correlators can be written as
\be
\label{Gdef}
\langle T\phi_{a}(x)\phi_{b}(y)\rangle=\delta_{ab}\,G(x,y).
\ee
As the $n_{k}$ grow, eventually the back-reaction from the quartic term will become important and the evolution will deviate from the quadratic approximation.
As we will see below (section (\ref{results})), this happens around the time when ($t=x^{0}$)
\be
\label{stopest}
M^2(t) = -\mu^2 +
\lambda\frac{N+2}{6N}\left[G(x,x)-G(0,0)\right]
\simeq 0.
\ee
Since
\begin{align}
G(x,x)-G(0,0) \simeq \intveck \frac{n_{k}(t)}{\omega_{k}(t)},
\end{align}
we expect the occupation numbers to be non-perturbatively large
\begin{align}
\label{nonpert}
n_{k}\sim 1/\lambda,
\end{align}
which suggests that a simple loop expansion in powers of the coupling
is unreliable for tachyonic preheating.

\section{Broken phase and finite volume}
\label{brokvol}
After the initial tachyonic preheating the system will go to
equilibrium and thermalize. To estimate the final temperature $T$,
consider a simple approximation in which the initial energy
density, $V_0$, is transfered to a free gas of $N-1$ massless
Goldstone bosons and one massive radial mode of mass $m
=\sqrt{2}\, \mu$. This approximation is expected to be reasonable
for weak coupling, provided $T$ is far from the critical temperature
$T_c$ at which the transition between the broken and
symmetric phase occurs. Using $V_0 = 3N\mu^4/2\lm$, the temperature
$T$ is then given by
\be
\label{Test}
\frac{3N\mu^4}{2\lm} =
\intveck \left(\frac{(N-1)k}{e^{k/T}-1} +
\frac{\sqrt{2\mu^2 + k^2}}{e^{\sqrt{2\mu^2 + k^2}/T}-1}\right).
\ee
For $N=4$, $\lm=6$ this gives $T=0.95\, \mu$ (the large $N$ limit would
give $0.93\, \mu$). On the other hand, the transition temperature
$T_c$ can be estimated
in a simple one-loop approximation \cite{Jansen:1989gd},
\be
T_c = \left(\frac{36 N}{(N+2)\lm}\right)^{1/2} m,
\ee
which gives $T_c = 2.83\, \mu$, for $N=4$ and $\lm=6$.
This also agrees well with Monte Carlo results \cite{Jansen:1989gd}.
Since the estimated final $T\ll T_c$,
we expect the system to end up in the broken-symmetry phase.

Our simulations are performed in a periodic box of size $L^3$. It is
well-known that quantum tunneling effects prevent
spontaneous symmetry breaking in finite volume. Hence,
$\langle \ph_a\rangle = 0$, but this does not mean that
physical effects of symmetry breaking cannot manifest themselves
in finite volume. In fact, the Goldstone bosons may cause relatively strong
finite size effects and slow thermalization.

The standard way to define $\langle \ph_a\rangle$ is to add a symmetry
breaking term to the action,
\be
\Delta S = \intx \ep\, n_a \ph_a,
\ee
which pulls $\langle \ph_a\rangle$ into the direction $n_a$,
and then evaluate
\be
v n_a = \lim_{\ep\to 0} \lim_{L\to\infty} \langle \ph_a\rangle.
\label{limlim}
\ee
Aided by finite-size scaling analysis this method has been used
fruitfully in Monte Carlo studies \cite{Hasenfratz:1990fu}.
The order of the limits in (\ref{limlim}) is important. However, reversing the
order (working at $\ep =0$), a lot can be learned from analysis at finite size
\cite{Hasenfratz:1989pk}.

A proper finite-size scaling analysis is outside
the scope of our present explorative study. Instead, we shall work at
reasonably large volumes, $L\mu >10$. As a guideline for interpretation
of the numerical results at relatively large times
(close to equilibrium), we take the tunneling into account by
simply averaging the infinite-volume broken-symmetry correlators
over all internal $O(4)$ directions:
\bea
\langle\ph_a\rangle &=& v\, n_a \to  0,
\\
\langle T\ph_a(x)\ph_b(y)\rangle
&=&
v^2 n_a n_b + (\dl_{ab}-n_a n_b) G^{\rm G}(x,y) + n_a n_b G^{\rm
H}(x,y)
\nonumber\\&\to&
\dl_{ab} G(x,y),
\label{O4av}
\\
G(x,y) &=&\frac{v^2}{N} + \frac{N-1}{N}\, G^{\rm G}(x,y) +
\frac{1}{N}\,G^{\rm H}(x,y),
\label{guide1}
\eea
where G and H denote the Goldstone and Higgs contributions. Hence,
after having settled sufficient into equilibrium, the $O(N)$
symmetric Green function $G$ is expected to have both Goldstone and
Higgs contributions, and a zero-momentum mode expressing the
condensate.

%
%

\section{The 2PI effective action and equations of motion\label{sseceom}}
%
%
The 2PI effective action can be written in the form \cite{Cornwall:1974vz}:
\be
\label{2Piform}
\Gamma[\bar{\phi},G]=S[\bar{\phi}]-\frac{i}{2}\Tr \ln{G}+\frac{i}{2}\Tr G_{0}^{-1}G+\Phi[\bar{\phi},G],
\ee
with
\begin{align}
iG^{-1}_{0,\,ab}(x,y)=
\frac{\delta^{2}\,S[\bar{\phi}]}{\delta\bar{\phi}_{a}(x)\,\bar{\phi}_{b}(y)}=
\left(\partial^{2}_{x}\,
\delta_{ab}+\mu^2\,
\delta_{ab}-\frac{\lm}{6N}\,\bar\ph_c\bar\ph_c \,
\delta_{ab}
- \frac{\lm}{3N}\,\bar\ph_a\bar\ph_b
\right) \delta(x-y).
\end{align}

We use the Schwinger-Keldysh formalism,
in which the fields live on a contour $\mathcal{C}$
running from $t=0$ to some time $t_{\rm max}$ and then back to $t=0$.
Time integrations and time ordering are along this contour.
At the two times $t_{\pm} = 0$ there are functional integrations
implementing the density matrix that specifies the initial state.
The functional $\Phi[\bar{\phi},G]$ is
a sum of 2PI skeleton diagrams with bare vertices and full propagators
corresponding to the initial state
, see figure \ref{figPhi1}.
\FIGURE{
\epsfig{file=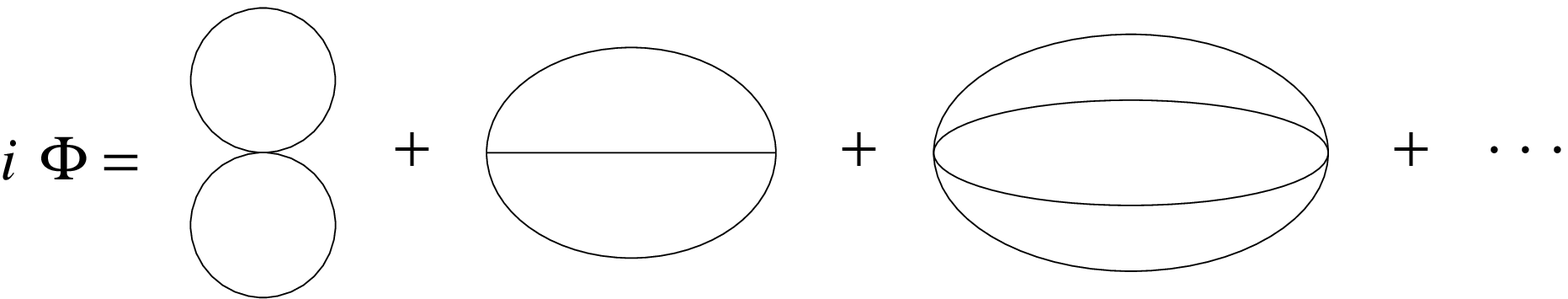,width=10cm,clip}
\caption{Expansion of $\Phi$ in terms of diagrams.
The lines denote the full $G$, the vertices denote the bare vertex
functions
$i\dl^n S[\bar\ph]/\dl\bar\ph_1 \cdots \dl\bar\ph_n$, $n=3,4$.
Symmetry factors are not indicated.}
\label{figPhi1}
}
%

The full propagator is a stationary point of the effective action. Taking the functional derivative
gives the equation determining $G^{ab}$,
\be
\frac{\delta \Gamma[\bar{\phi},G]}{\delta G^{ab}(x,y)}=0
\to
G^{-1}_{ab}(x,y)=G_{0,ab}^{-1}(x,y)+i\Sigma_{ab}(x,y),
\label{eqG1}
\ee
with $\Sg$ the self-energy,
\be
\Sg_{ab}(x,y) =
-2\frac{\delta \Phi[\bar{\phi},G]}{\delta G^{ab}(x,y)}.
\label{sigmadef}
\ee
Multiplying (\ref{eqG1}) by $G^{bc}(y,z)$ turns it into
an integro-differential equation,
\be
\label{eqG2}
\delta_{ac}\delta(x,z)=\int_{\mathcal{C}} dt'\int d^{3}y\,\left[
G^{-1}_{0,\, ab}(x,y)+i\Sigma_{ab}(x,y)\right]\,G^{bc}(y,z).
\ee
Together with the equation for $\bar\ph$,
which follows from $\dl\Gm[\bar\ph,G]/\dl\bar\ph^a = 0$,
this determines $G(y,z)$ in a self-consistent way, since the self-energy is itself a functional of $G$, eq. (\ref{sigmadef}).
Since $\bar\ph$ will be identically zero because of the $O(N)$ symmetry
of the initial state, we now specialize to this case.

The delta function in (\ref{eqG2}) for $G$ can be taken care of by introducing
$G^>$ and $G^<$ according to
\be
G^{ab}(x,y)\equiv \delta_{ab}\left[G^{>}(x,y)\,\theta_{\mathcal{C}}(t-t')+G^{<}(x,y)\,\theta_{\mathcal{C}}(t'-t)\right],
\ee
where the step functions $\theta_{\mathcal{C}}(t-t')$ refer to the time label along the contour $\mathcal{C}$.
The `statistical function' $F$ and the spectral function $\rh$
correspond to the real and imaginary parts of $G^{>}$ or $G^<$,
\be
G^{>}(x,y)=F(x,y)-i\rho(x,y)/2,\\ G^{<}(x,y)=F(x,y) +i\rho(x,y)/2.
\ee
In operator language they correspond to
\be
F(x,y)\delta_{ab}=\frac{1}{2}\langle\{\phi^{a}(x),\phi^{b}(y)\}\rangle,\qquad\rho(x,y)\delta_{ab}= i\langle[\phi^{a}(x),\phi^{b}(y)]\rangle.
\ee
The functions $F(x,y)$ and $\rho(x,y)$ are real and have the symmetry properties
\be
\label{symeq1}
F(x,y)=F(y,x),~~~~~\rho(x,y)=-\rho(y,x).
\ee
The canonical commutation relations imply
\be
\label{symeq2}
\partial_{0}\rho(x,x')|_{x^0=x^{\prime\,0}}=-\partial_{0'}\rho(x,x')|_{x^0=x^{\prime\,0}}=\delta({\bf x,x'}).
\ee
Equation (\ref{eqG2}) now leads to coupled integrodifferential
equations for $F$ and $\rh$, supplemented by initial conditions depending
on the initial density matrix. These equations preserve the properties eq. (\ref{symeq1}, \ref{symeq2}), and conserve energy for any choice of truncation  of $\Phi[\bar{\phi},G]$.

One would like to know if truncations of $\Phi$ lead to solutions for $G$ that
describe massless Goldstone bosons when the system is in the broken-symmetry
phase\footnote{Such Goldstone bosons are still present in our case of zero
mean field, cf.\ section \ref{brokvol}.}. Generically this is not the case
\cite{Aarts:2002dj,vanHees:2002bv}. 
However, one can introduce the 1PI effective action
\be
\tilde \Gm[\bar\ph] \equiv \Gm[\bar\ph,G[\bar\ph]],
\ee
with $G[\bar\ph]$ the solution of (\ref{eqG1}) for given mean field $\bar\ph$,
and
define a Green function $\tilde G$ (called `external' in \cite{vanHees:2002bv})
in the usual way from $\tilde\Gm$:
\be
i \tilde G^{-1}_{ab}(x,y) =
\frac{\dl^2\tilde\Gm[\bar\ph]}{\dl\bar\ph^a(x)\, \dl\bar\ph^b(y)}.
\ee
Then the usual derivation of Goldstone's theorem in terms of $\tilde\Gm$ applies
to $\tilde G^{-1}$, it has $N-1$ zero eigenvalues at zero momentum 
 \cite{Aarts:2002dj,vanHees:2002bv}.
Without approximation the Green functions $\tilde G$ and $G$ coincide, but this is no longer true for a truncated $\Phi$.
The difference between $G$ and $\tilde G$ depends on the functional derivative
$\dl G[\bar\ph]/\dl\bar\ph$ \cite{vanHees:2002bv}.
This is a three-point function (given by the solution of a non-trivial
integral equation) that
vanishes in our case of zero mean field. Therefore, we do expect $G$ to describe massless Goldstone bosons.

At first we tried a truncation of the effective action based on
the loop expansion, by keeping the diagrams shown in 
figure \ref{figPhi1}, for $N=1$. However, this lead to severe
instabilities before the transition was completed, for weak and
also for stronger couplings, which we were unable to cure. We
interpret this as a breakdown of an expansion based on the
smallness of $\lm$, because of the non-perturbatively large
occupation numbers expected during the instability (eq.
(\ref{nonpert})). We then turned to an expansion of the 2PI
effective action, not in powers of the coupling
$\lambda$, but
in terms of $1/N$.
Following \cite{Berges:2001fi} we expand $\Phi$
to Next-to-Leading Order (NLO) in $1/N$.
The explicit form of the resulting equations can be found in
\cite{Berges:2001fi} (see \cite{Aarts:2002dj} for the case of non-zero mean field).

\section{Numerical implementation\label{LargeN}}

We discretize the $O(N)$ model on a space-time lattice with the action
\begin{align}
\label{latact}
S_{\rm lat}=a_s^3 a_t\sum_{{\bf x},t} \left(
\frac{\left(\partial_{t}\phi_{a}(t,{\bf x})\right)^{2}}{2}
-\frac{\sum_{i}\left(\partial_{i}\phi_{a}(t,{\bf x})\right)^{2}}{2}
-\frac{\mu_{0}^{2}}{2}\,\phi_{a}^{2}(t,{\bf x})
-\frac{\lambda_0}{24N}
\left(\phi_{a}^2(t,{\bf x})\right)^{2}+V_{0}\right).
\end{align}
Here the lattice spacings are $a_s$ and $a_t$ in the spatial and time
directions,
$\partial_{t}\phi(t,{\bf x})=(\phi(t+a_t,{\bf x})-\phi(t,{\bf x}))/a_t$,
and similarly for spatial derivatives.
We use periodic boundary conditions, with volume is $L^{3}=(n_{s}a_{s})^{3}$.
In the following
we shall use lattice units, $a_s=1$ and write $dt$
for the dimensionless time-step ($dt = a_t/a_s$).
The coupling $\lm_0$ and mass parameter $\mu^{2}_{0}$ are {\em bare}
parameters to be determined below.
At tree level, $\lm_0=\lm$ and $\mu_{0}^{2}=-\mu^{2}$ in the broken phase.
The lattice version of the squared momentum is given by,
\begin{align}
k^{2}_{\rm lat}=\sum_{i=1}^3(2-2\cos{k_{i}}),\qquad
k_{i}=\frac{2\pi n_{i}}{n_s},
~n_{i}=-\frac{n_s}{2}+1,\cdots,\frac{n_s}{2}
\end{align}
($n_s$ is chosen even). Plotting data as a function of the lattice
momenta rather than continuum $k^{2}$ corrects for most of the
lattice artifacts.

As explained in \cite{Berges:2001fi}, truncating the $1/N$ expansion at NLO and specializing to a homogeneous system ($G(x,y)=G(t,t',{\bf x}-{\bf y})$)
with zero mean field, leads to the equations of motion:
\begin{align}
\label{eomFN}
\partial_{t}\partial_{t}'F(t,t',{\bf x})=&\,\partial_{i}\partial_{i}'F(t,t',{\bf x})-M^{2}(t)F(t,t',{\bf x})\nonumber\\
&+\sum_{t''=0}^{t}\,dt \,\sum_{\bf z}\Sigma_{\rho}(t,t'',{\bf z})F(t'',t',{\bf x-z})\nonumber\\
&-\sum_{t''=0}^{t'}\,dt \,\sum_{\bf z}\Sigma_{F}(t,t'',{\bf z})\rho(t'',t',{\bf x-z}),\\
\label{eomrhoN}
\partial_{t}\partial_{t}'\rho(t,t',{\bf x})=&\,\partial_{i}\partial_{i}'\rho(t,t',{\bf x})-M^{2}(t)\rho(t,t',{\bf x})\nonumber\\
&+\sum_{t''=t'}^{t}\,dt \,\sum_{\bf z}\Sigma_{\rho}(t,t'',{\bf z})\rho(t'',t',{\bf x-z}),
\end{align}
with
$\partial^{'}_{t}\phi(t,{\bf x})=(\phi(t,{\bf x})-\phi(t-dt,{\bf x}))/dt$,
and similar for $\partial'_i$, and with\footnote{When practical,
we shall use a notation in which, for instance, $\rho(t,t',{\bf
x})F(t,t',{\bf x})=(\rho F)(t,t',{\bf x})$.}
\begin{align}
\label{MdefN}
M^{2}(t)=&\,\mu_{0}^{2}+\lambda_0\frac{N+2}{6\,N}F(t,t,{\bf 0}),\\
\Sigma_{F}(t,t'',{\bf z})=&\,\frac{\lambda_0}{3N}\left(F\, I_{F}-\rho \,I_{\rho}/4\right)(t,t'',{\bf z}),\\
\Sigma_{\rho}(t,t'',{\bf z})=&\,\frac{\lambda_0}{3N}\left(F\, I_{\rho}+\rho \,I_{F}\right)(t,t'',{\bf z}).
\end{align}
The functions $I_{F}$ and $I_{\rho}$
represent the summation of bubble diagrams included at NLO \cite{Berges:2001fi}
\begin{align}
\label{eomIF}
I_{F}(t,t',{\bf x})=&\,\frac{\lambda_0}{6}\left(F^{2}-\rho^{2}/4\right)(t,t',{\bf x})\nonumber\\
&+\frac{\lambda_0}{3}\sum_{t''=0}^{t}dt\sum_{\bf z}\,I_{F}(t,t'',{\bf z})\left(F\rho\right)(t'',t',{\bf x-z})\nonumber\\
&- \frac{\lambda_0}{6}\sum_{t''=0}^{t'}dt\sum_{\bf z}\,I_{\rho}(t,t'',{\bf z})\left(F^{2}-\rho^{2}/4\right)(t'',t',{\bf x-z}),\\
\label{eomIR}
I_{\rho}(t,t',{\bf x})=&\,\frac{\lambda_0}{3}\left(F\rho\right)(t,t',{\bf x})\nonumber\\
&-\frac{\lambda_0}{3}\sum_{t''=t'}^{t}dt\sum_{\bf z}\,I_{\rho}(t,t'',{\bf z})\left(F\rho\right)(t'',t',{\bf x-z}).
\end{align}
Note that $\Sg_F$ and $\Sg_\rh$ contain an overall factor $\lm_0^2$,
they would start at order $\lm_0^2$ upon solving the above equations by
perturbation in $\lm_0$. The contribution of order $\lm_0$ (coming from the first
diagram in figure \ref{figPhi1}) is purely local (non-zero only for $t=t''$)
and contained in $M^2$ defined in (\ref{MdefN}).
During the numerical evolution we have to solve for $I_{F}$ and $I_{\rho}$ parallel to solving for $F$ and $\rho$. Although eqs. (\ref{eomIF}, \ref{eomIR}) look implicit, by doing things in the right order and taking advantage of the symmetries
\be
I_{F}(t,t',{\bf x})=I_{F}(t',t,{\bf x}),~~~I_{\rho}(t,t',{\bf x})=-I_{\rho}(t',t,{\bf x}),~~~I_{\rho}(t,t,{\bf x})=0,
\ee
they can be solved explicitly. The energy functional is given by
\begin{align}
\label{energy2PIN}
\langle T^{00}(t,{\bf 0})\rangle =&\,\frac{\partial_{t}\partial_{t'}F(t,t',{\bf 0})}{2}|_{t=t'}
+\frac{\sum_i\partial_i\partial'_i F(t,t,{\bf x})}{2}|_{{\bf x}={\bf 0}}
+\frac{\mu_{0}^{2}}{2}F(t,t,{\bf 0})+\frac{\lambda_0(N+2)}{24N}\,F^{2}(t,t,{\bf 0})\nonumber\\
&-\frac{\lambda_0}{12\,N}\sum_{t''=0}^{t}dt\sum_{\bf z}\,\left(2\,\rho \,F\, I_{F}+(F^{2}-\rho^{2}/4)\,I_{\rho}\right)(t,t'',{\bf z})
- \mbox{sub.},
\end{align}
where `sub.' is a subtraction. Ideally, this subtraction is such that the
energy density is zero in the vacuum (the zero-temperature ground state), but
its evaluation requires a non-trivial minimization. In practise we set the
energy density to $V_0$, for the initial state (\ref{init1}--\ref{init3}).

\subsection{Hartree  approximation}
\label{Hartree}
In addition to the NLO-1/N approximation, we
study
the Hartree approximation, which results from
taking into account only the first diagram in 
figure \ref{figPhi1}. This means
leaving out all time integrals, and the $I_{F,\rho}$, keeping only the first line of eqs.\ (\ref{eomFN}) and (\ref{eomrhoN}), with $M^{2}$ defined by eq.\
(\ref{MdefN}).
Notice that this is {\it not} the Leading-Order (LO) approximation in 1/N, since the coefficient of the local self-energy is $\lambda_0\frac{N+2}{6N}$ as opposed to $\frac{\lambda_0}{6}$ for LO-1/N.
Neither of the two includes non-trivial scattering in the dynamics
and they give qualitatively the same results. We
use Hartree dynamics only for comparison.

%
%

\subsection{Initial conditions\label{ssecinitTFT}}

We restrict ourselves to initial conditions
for which the density matrix is gaussian. This assumption is
equivalent to the initial state being completely described in
terms of (one- and) two-point functions \cite{Cooper:1997ii}.
Consequently, an initial condition is a choice of initial $F$'s,
$\rho$'s and $\bar{\phi}$ (which in this case is
zero). We need to specify
\begin{align}
\frac{1}{2}\langle\left\{\phi^{a}(0,{\bf x}),\phi^{b}(0,{\bf 0})\right\}\rangle&= \,\delta_{ab}\,F(t,t',{\bf x})|_{t=t'=0},\\
\frac{1}{2}\langle\left\{\pi^{a}(0,{\bf x}),\phi^{b}(0,{\bf 0})\right\}\rangle&=\,\delta_{ab}\,\partial_{t}F(t,t',{\bf x})|_{t=t'=0},\\
\frac{1}{2}\langle\left\{\pi^{a}(0,{\bf x}),\pi^{b}(0,{\bf 0})\right\}\rangle&=\,\delta_{ab}\,\partial_{t}\partial_{t'}F(t,t',{\bf x})|_{t=t'=0},
\end{align}
in addition to
\begin{align}
i\langle\left[\phi^{a}(0,{\bf x}),\phi^{b}(0,{\bf x})\right]\rangle&= \delta_{ab}\rho(0,0,{\bf x})=0,\\
i\langle[\pi^{a}(0,{\bf x}),\phi^{b}(0,{\bf 0})]\rangle&=\,\delta_{ab}\,\partial_{t}\rho(t,t',{\bf x})|_{t=t'=0}=\,\delta_{ab}\,\delta_{{\bf x},{\bf 0}},\\
\end{align}
where we used $G(x,y)\equiv G(t,t',{\bf x-y})$ for a homogeneous system.
This also allows us to go to momentum space and
define\footnote{The Fourier transforms are given by
$F_\veck(t,t') = L^{-3/2}\sum_{\bf x}F(t,t',{\bf x})\,e^{-i{\bf kx}}$,
etc.}
\begin{align}
\label{nkomk}
\frac{1}{2}\langle\left\{\phi^{a}_{\bf k}(0),\phi^{b}_{\bf -k}(0)\right\}\rangle
=&\,\delta_{ab}F_\veck(0,0)
\equiv\frac{n^{\rm init}_{k}+1/2}{\omega^{\rm init}_{k}}\,\delta_{ab},\\
\frac{1}{2}\langle\left\{\pi^{a}_{\bf k}(0),\phi^{b}_{\bf -k}(0)\right\}\rangle
=&\,\delta_{ab}\,\partial_{t}F_\veck(t,t')|_{t=t'=0}
\equiv \tilde{n}^{\rm init}_{k}\,\delta_{ab},\\
\frac{1}{2}\langle\left\{\pi^{a}_{\bf k}(0),\pi^{b}_{\bf -k}(0)\right\}\rangle
=&\,\delta_{ab}\,\partial_{t}\partial_{t'}F_\veck(t,t')|_{t=t'=0}
\equiv (n^{\rm init}_{k}+1/2)\,\omega^{\rm init}_{k}\,\delta_{ab},
\end{align}
given a choice of ``occupation number distributions'' $n^{\rm init}_{k}$,
$\tilde{n}^{\rm init}_{k}$
and dispersion relation $\omega^{\rm init}_{k}$. We
specialize to the case $\tilde{n}^{\rm init}_{k}=0$, noting that
in equilibrium  this correlator is zero. As mentioned in section \ref{sectacpre}, we assume that the system is initially in the state corresponding to the ground state before the quench, which we approximate by the free-field vacuum
in the symmetric-phase,
\begin{align}
\label{initdef1}
n^{\rm init}_{k}=0, \qquad
\omega^{2,\rm init}_{k}=\mu^2+k^{2}_{\rm lat}.
\end{align}
Notice that we use the renormalized quantity $\mu^2$ in the initial condition,
in accordance with section \ref{sectacpre}.
Also, since this initial state is a free-field vacuum,
it is not the true vacuum of
our interacting system in the symmetric phase.

%
%

\subsection{Choice of bare parameters \label{renorm1}}

The renormalizability of $\Phi$-derivable approximations was studied in 
\cite{vanHees:2001ik,VanHees:2001pf,Blaizot:2003an,Blaizot:2003br,Berges:2004hn}.
We assume the NLO-1/N approximation to be renormalizable, 
in the sense that we can
 choose the coefficients in the discretized equations
(\ref{eomFN}--\ref{eomIR}) in such a way that the lattice spacing
becomes exceedingly small
\footnote{This may imply a more elaborate
set of parameters than just $\mu_0^2$ and $\lm_0$. At worst, an independent 
parameter may be needed for every contribution to the truncated $\Phi$ 
(e.g.\ every diagram), since there is no symmetry relating various
contributions. 
 In practise this is not a problem 
for sufficiently accurate truncations \cite{Berges:2004hn}.} 
-- loosely called `the continuum
limit'.\footnote{Such terminology ignores triviality.}
We will not study lattice-spacing dependence but do want to choose
the parameters $\mu_0^2$ and $\lm_0$ such that the relevant length scale in our simulation is larger than the discretization scale, i.e.\
$\mu^{-1} > a_s$.

For this purpose we implement renormalization at one loop order. At weak coupling,
mass renormalization is important, but not coupling renormalization. Two-loop effects are also small, and the bare mass $\mu_0^2$ in the equation of motion is therefore estimated in one-loop renormalized
perturbation theory as
\bea
\label{masscounter}
\mu_{0}^{2}&=&\pm \mu^2-\lambda\frac{N+2}{6N}\, c_{1}(a_s\mu)\,
\frac{1}{a_s^2},
\\
c_1(a_s\mu) &=& -i \int_{-\pi/a_s}^{\pi/a_s}\frac{d^3 k}{(2\pi)^3}\,
\int \frac{dk_0}{2\pi}\,
\frac{a_s^2}{\mu^2 + a_s^{-2} k_{\rm lat}^2 - k_0^2 -i\ep}
=
\int_{-\pi}^{\pi}\frac{d^3 k}{(2\pi)^3}\,
\frac{1}{2\sqrt{a_s^2\mu^2 + k_{\rm lat}^2}}
\\
&=&
C_0  + C_2\, a_s^2\mu^2 + \frac{1}{16\pi^2}\, a_s^2\mu^2 \ln(a_s^2\mu^2)
+ O(a_s^4\mu^4),
\eea
in the continuous-time and infinite-volume limit. In
(\ref{masscounter}) the plus (minus) sign refers to the symmetric
(broken) phase. The one-loop self-energy refers to the symmetric
phase, the broken phase should give the same divergent
contribution as $a_s\to 0$ and a small difference in the finite
contribution.
An idea of the size of the coefficients may be obtained from a calculation
on a symmetric ($a_t = a_s$) lattice in imaginary time \cite{Sm02},
\be
C_0 = 0.155\ldots,
\quad
C_2 = 0.0303\ldots ,
\quad a_t=a_s.
\ee
Similarly, the bare coupling can be estimated by the one-loop expression
\cite{Sm02}
\be
\lm_0 = \lm - \lm^2\frac{N+8}{6N}\left(\frac{1}{16\pi^2}\, \ln(a_s^2 \mu^2)
+ \frac{1}{16\pi^2} - C_2\right) + O(a_s^2\mu^2)
\label{couplingcounter}
\ee
(where $\lm$ is defined as the value of the four-point vertex
function at vanishing external momenta). For the maximum coupling
that we use ($\lm = 6$) and lattice spacing $0.5< a_s\mu < 1$, the
difference between $\lm_0$ and $\lm$ is less than 10\%. In
practise we simply choose $\lm_0$ as if it were the renormalized
coupling.
In our simulations we use the bare coupling in
(\ref{masscounter}), i.e.\ on a $n_{s}^{3}$-site lattice, in
lattice units,
\begin{align}
\label{masscounter2}
\mu_{0}^{2}=&-\mu^2-\lambda_0\frac{N+2}{6N}\,c_{1}(\mu),
\\
\label{c1def}
c_{1}(\mu)=&
\frac{1}{n_s^{3}}\sum_{{\bf k}}\frac{1}{2\sqrt{\mu^2 + k_{\rm lat}^2}}
\simeq 0.21\, .
\end{align}
Given the input parameters $\mu_0^2$ and $\lm_0$, the output
physics is of course not known precisely, it is determined by the
$\Phi$-derived equations. For example, the mass of the (unstable)
Higgs particle will differ from the tree graph value $m =
\sqrt{2}\, \mu$. Similarly, the renormalized coupling will not
satisfy the relation (\ref{couplingcounter}) precisely ($\lm$ may
also be defined in terms of the vev by $v^2 = 3 N m^2/\lm$ where
the vev may be identified from the low-energy scattering amplitude
of the Goldstone bosons,
cf.\ (\ref{GBscat})). But the differences are expected to be
small, and they can in principle be determined numerically.

In our out-of-equilibrium study there is another dimension-full scale: the mass
$\mu$ in the initial conditions (\ref{initdef1}). It is
the renormalized mass in the symmetric phase, and it is this mass
that we use in (\ref{masscounter2}) for choosing $\mu_0^2$. We
have to keep in mind that the gaussian approximation in the
initial conditions and the suddenness implied by the quench may
lead to divergencies in the pressure near $t=0$, in the continuum
limit, see e.g.\ \cite{Baacke:1997zz}. These may be dealt with by
adjusting the initial density matrix, but it would be better to
avoid the instantaneous quench and use a more realistic model of
the transition. Such divergencies do not appear to be important in
our simulations at finite lattice spacing.

%
%

\subsection{Observables\label{ssecobservables}}

We follow the evolution of $F(t,t,{\bf 0})$ and the energy, eq.\
(\ref{energy2PIN}), and calculate occupation numbers
$n_{k}$ and frequencies $\omega_{k}$ through the inverted version
of eq.\ (\ref{nkomk}),
\begin{align}
\label{nkomdef}
n_{k}+1/2=&c_{k}\sqrt{\partial_{t}\partial_{t'}F_\veck(t,t')|_{t=t'}
F_\veck(t,t)},\\
\omega_{k}=&\sqrt{\partial_{t}\partial_{t'}F_\veck(t,t')|_{t=t'}/
F_\veck(t,t)}.
\label{nkomdef2}
\end{align}
The quantities $n_{k}$ and $\omega_{k}$ coincide with the true particle occupation numbers and frequencies when applied to a free field in equilibrium, and have proven to be very useful out-of-equilibrium in interacting theories as well \cite{SaSm00a,SaSm00b,Salle:2002fu,Skullerud:2003ki}. The correction $c_{k}$ takes care of errors associated with the time discretization on the lattice. It is given by
\cite{Salle:2002fu}
\be
c_{k}=\sqrt{1-\frac{1}{4}dt^{2}\omega_{k}^{2}}.
\ee

In the present case we argued in section \ref{brokvol} that the
two-point function
$G$ receives contributions from both the Higgs and the Goldstone
modes, so in this respect the $n_k$ and $\om_k$ are
{\em compound} observables.

After some initial transients, the system will approach thermal
equilibrium, although this may take a long time because of
critical slowing down: the scattering of the Goldstone bosons
becomes suppressed for momenta smaller than the Higgs mass, cf.\
(\ref{GBscat}). The approximate expression (\ref{guide1}) for
$G$ in terms of $G^{\rm H}$ and $G^{\rm G}$ provides a guide to
what form we may expect for the $n_k$ and
$\om_k$ defined in (\ref{nkomdef}):
\bea
F_\veck(t,t) &\approx& \frac{v^2(t) L^3}{N}\, \dl_{\veck,{\bf 0}} +
\frac{N-1}{N}\,\frac{n_k^{\rm G}(t) + 1/2}{\om_k^{\rm G}(t)} +
\frac{1}{N}\, \frac{n_k^{\rm H}(t) + 1/2}{\om_k^{\rm H}(t)},
\label{Fguide}\\
\partial_{t}\partial_{t'}F_\veck(t,t')|_{t=t'} &\approx&
\frac{N-1}{N}\,\left(n_k^{\rm G}(t) + 1/2\right)\om_k^{\rm G}(t) +
\frac{1}{N}\, \left(n_k^{\rm H}(t) + 1/2\right)\om_k^{\rm H}(t),
\label{Fddguide}
\eea
where, assuming that the individual Higgs and Goldstone modes have
thermalized to a Bose-Einstein (BE) distribution,
\be
\om_k^{\rm H} = \sqrt{m^2 + k_{\rm lat}^2},
\quad
\om_k^{\rm G} = k_{\rm lat},
\quad
n_k^{\rm H} = \frac{1}{e^{\om_k^{\rm H}/T}-1},
\quad
n_k^{\rm G} = \frac{1}{e^{\om_k^{\rm G}/T}-1},
\label{BE}
\ee
and $v$ and $m$ are to be interpreted as the condensate and
effective Higgs mass at temperature $T$. We have allowed for a time dependent $v(t)$, although in (\ref{Fddguide}) we have neglected its time derivatives. Close to equilibrium we expect the change in $v$ to be very slow ($\dot{v}/v\ll\mu$, see figure \ref{Cond} below). At low momenta ($k< m$)
the Goldstone bosons will clearly dominate. Note that in finite
volume the condensate appears only in the zero mode of $F$, and
there is no zero mode in the Goldstone contribution $G^{\rm G}$:
it is represented by the global rotations making up the $O(4)$
average in (\ref{O4av}).

\FIGURE{
\epsfig{file=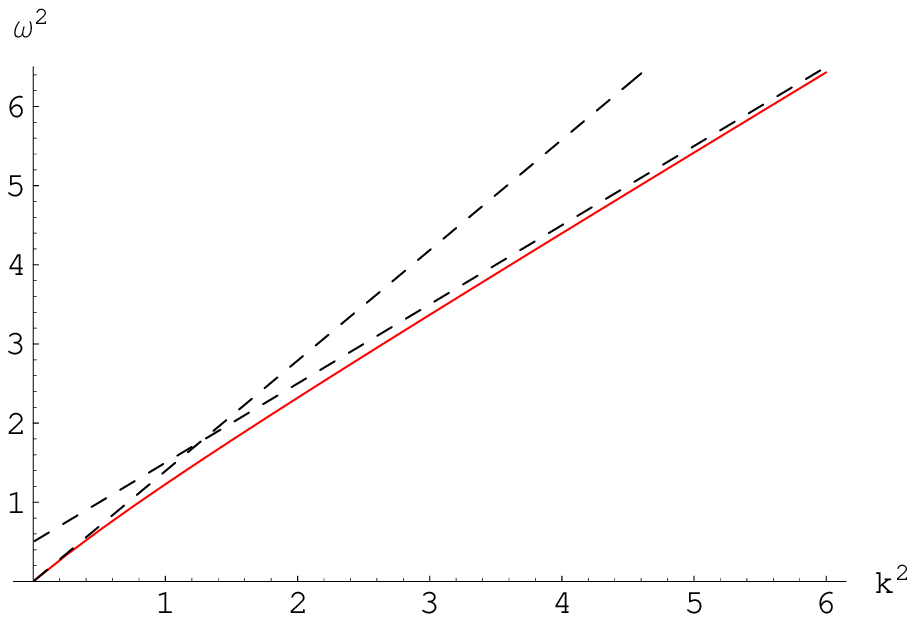,width=7cm,clip}
\epsfig{file=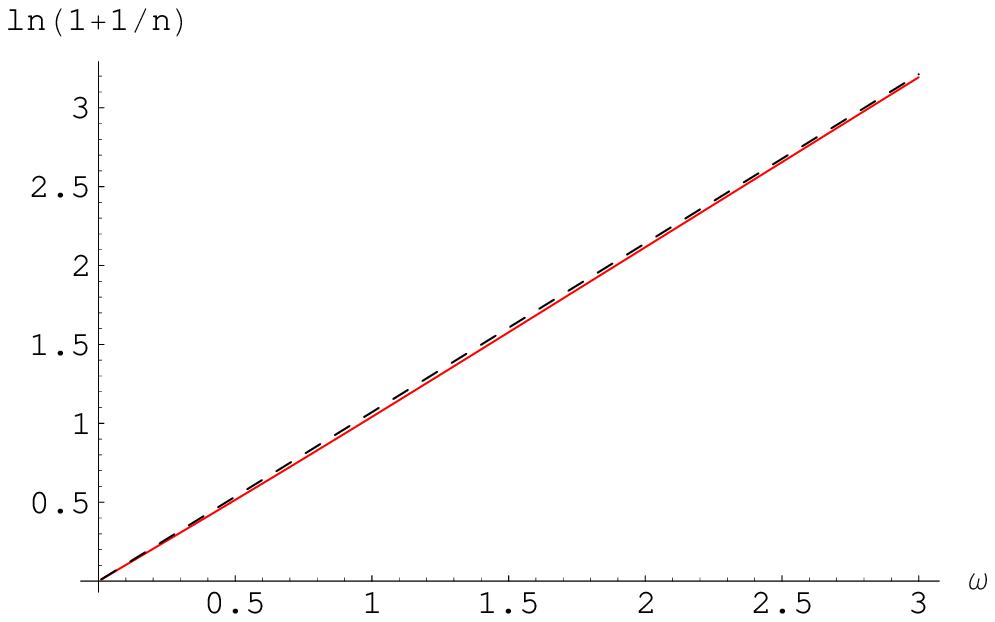,width=7cm,clip}
\caption{Left: Compound frequency spectrum from
(\ref{nkomdef})--(\ref{BE}),
$\om_k^2$ versus $k^2$, for $N=4$, $T=0.95\, \mu$,
$m=\sqrt{2}\, \mu$, in units $\mu=1$.
The dashed lines $\om^2 = 1.395\, k^2$ and $\om^2 = 0.5\mu^2 + k^2$
are the asymptotes (\ref{omas0}) and
(\ref{omasinfty}). Right: Compound particle number,
$\ln(1+1/n_k)$ versus $\om_k$.
The dashed line represents $\om/T$.
\label{omnguide} }
}

In figure
\ref{omnguide} (left) we have plotted the resulting $\om_k$ at the
expected temperature
$T=0.95\, \mu$ found in section \ref{brokvol} for $N=4$, $\lm=6$.
The straight lines correspond to the asymptotes
\bea
\om_k^2 &=& \left[1+ \frac{m\left(n_0^{\rm H}+1/2\right)}
{T(N-1)}\right] k^2 + O(k^4),
\quad k\to 0,
\label{omas0}
\\
&=& k^2 + \frac{m^2}{N} + O(k^{-2}),
\quad k^2\gg m^2, T^2.
\label{omasinfty}
\eea
This latter formula may provide an estimate of the effective Higgs
mass, although the emphasis on the large momentum region is
questionable. A better way to determine $m$ should be to find the
position of the peaks in the Fourier transform of the spectral
function, as done in 1+1 dimensions in \cite{Aarts:2001qa}. Figure
\ref{omnguide} (right) shows the compound particle number from
(\ref{nkomdef})--(\ref{BE}) for the same parameters
as for $\om_k$ in the left-figure. Plotted is $\ln(1+1/n_k)$
versus
$\om_k$, which would be a straight line with slope $1/T$ if $n_k$
were a Bose-Einstein distribution with zero chemical potential.
The deviation from BE is hardly visible, which suggests that such
a plot is also useful for judging the approach to equilibrium,
even for the compound particle number. For a BE with a non-zero chemical potential $\mu_{c}$, we would see an intercept with the $\omega_{k}$-axis equal to $\mu_{c}$. For the compound particle number this also applies with the understanding that the intercept gives the chemical potential of the {\it Goldstone} particles, $\mu_{\rm ch}^{\rm G}$.

For sufficiently large
volume the condensate follows from the zero mode of $F$,
\be
\label{condest}
\frac{v^2}{N}
\simeq
\frac{F_{\bf 0}(t,t)}{L^3},
\ee
which has been found to give good results compared to a more rigorous
procedure based on (\ref{limlim}) and finite-size scaling \cite{Hasenfratz:1990fu}.
The relative correction
\be
\frac{n^{\rm H}_0 +1/2}{m v^2 L^3} \approx
\frac{\lm (n_0^{\rm H}+1/2)}{6\sqrt{2}\, N (\mu L)^3}
\ee
(which may also be estimated from
$\partial_t\partial_{t'}F_{\bf 0}(t,t')_{t=t'}/F_{\bf 0}(t,t) m^2$) is tiny:
for $\mu L = 10$ and $\lm n_0^{\rm H}$ of order 100 it is only
of order $10^{-2}/N$.

In the classical approximation we found
\cite{Smit:2002yg,Tranberg:2003gi,Skullerud:2003ki}
the observable $N^{-1}\langle\ph_a(x)\ph_a(x)\rangle$ useful
for monitoring the tachyonic transition, and we have used it here as well.
Its quantum analog $F(x,x) = F(t,t,{\bf 0})$ is divergent in the continuum
limit and needs a subtraction. We have taken this to be its initial value
$F(0,0) = F(0,0,{\bf 0}) = c_1(\mu)$ ($ \simeq 0.21$ in lattice units
cf.\ (\ref{nkomk}), (\ref{initdef1}), (\ref{c1def})).
The difference $F(x,x)-F(0,0)$ then vanishes at time zero
and at late times (\ref{Fguide}) indicates that
\bea
F(x,x)-F(0,0) &=& \frac{v^2}{N} +
\frac{1}{n_s^3}\sum_{\veck} \left(
\frac{N-1}{N}\,\frac{n_k^{\rm G}+1/2}{\om_k^{\rm G}}(1-\dl_{\veck,{\bf 0}}) +
\frac{1}{N}\, \frac{n_k^{\rm H}+1/2}{\om_k^{\rm H}}
-\frac{1}{2\sqrt{\mu^2 + k_{\rm lat}^2}}\right)
\nonumber\\
&\approx& \frac{v^2}{N} + \intveck\left(
\frac{N-1}{N}\,\frac{n_k^{\rm G}}{\om_k^{\rm G}} +
\frac{1}{N}\, \frac{n_k^{\rm H}}{\om_k^{\rm H}}
\right)
\nonumber\\&&\mbox{}
+ \half \intvecklat \left(
\frac{N-1}{N}\, \frac{1}{k_{\rm lat}} +
\frac{1}{N}\, \frac{1}{\sqrt{m^2 + k_{\rm lat}^2}}
- \frac{1}{\sqrt{\mu^2 + k_{\rm lat}^2}}\right),
\label{lastint}
\eea
where the second line refers to $a_s\neq 1$ units. The last
integral is logarithmically divergent as
$a_s\to 0$ (at two-loop accuracy there may be also quadratic divergencies).
A different subtraction, e.g.\ such that the result is finite in
the broken-symmetry vacuum state, may render it divergent as
$t\to 0$. This is an example of the artifacts introduced by the
suddenness of the quench. Nevertheless, for the lattice spacings
that we use these remaining divergencies are negligible. For
example, in lattice units the last integral in (\ref{lastint}) is
only 0.00922 for
$m^2 = 2\mu^2 = 1$.

We also compute the `memory kernels'
in the equations of motion
(\ref{eomFN})--(\ref{eomrhoN}), the self-energies
$\Sigma_{F\,\veck}(t,t')$ and
$\Sigma_{\rho\,\veck}(t,t')$, which can be compared with
perturbative estimates. For long runs, finite computer memory
requires us to ``cut off'' the self-energy (in terms of the
functions $I_{F,\rho}$), by keeping the memory kernels only some
finite range backwards in time (i.e.\ $\Sg(t,t'',\vecz) \to 0$,
$|t-t''| > t_{\rm cut}$). The size of the self-energies will help
us determine whether we have cut off late enough for the discarded
memory integrals to be negligible for the dynamics. A too early
cut-off also shows up as a non-conservation of energy. In all the
runs presented here, the energy is conserved to within two
percent.
At early times
its fluctuations are due to time discretization errors. At very
late times, the effect of cutting off the memory kernel can add up
to a very slow drift in the energy, but still within the two
percent band. Keeping the whole kernel (say, for lattices of
smaller spatial extend) there is no such drift, and we have
checked that later cut-offs make the drift smaller. Our choices of
memory cut-offs reflect a compromise between exact energy
conservation, lattice sizes, computer memory and especially CPU
time, and the wish to study reasonably long time evolution.

%
%

\section{The classical approximation}

In the classical approximation, a set of classical initial
realizations is evolved using the classical equations of motion.
These realizations are drawn from an ensemble of initial
conditions that reproduce the initial quantum correlators in a
classical approximation. In the case of tachyonic preheating we do
not have to wait for the correlators to become classical if the
couplings are weak -- we can go back all the way to time zero and
reproduce eqs.\ (\ref{eqC1}), (\ref{eqC2}), (\ref{eqC3}) at
$t=0$,
\be
\label{quanthalf}
C_k^{\ph\ph}(0) = \frac{1}{2\sqrt{\mu^2 + k^2}},
\qquad C_k^{\pi\pi}(0)=\frac{\sqrt{\mu^2 + k^2}}{2},
\quad
k < \mu.
\ee
See \cite{Smit:2002yg} for a more detailed justification of this
procedure. Note that the stable modes $k>\mu$ are initialized to
zero. In section \ref{whichhalf} we comment on the case where all
the modes are initialized according to the initial quantum
correlators, including the modes
$k>\mu$. Observables are then averaged over the sample, giving an
approximation of the quantum observable,
\begin{align}
\langle\mathcal{O}\rangle_{\rm quant.}\simeq\langle\mathcal{O}\rangle_{\rm class. ensemble}
\end{align}

To compare with the $\Phi$ derived approximation,
we perform classical simulations starting from the same lattice Lagrangian as in the quantum case, eq. (\ref{latact}). The classical equation of motion reads
\begin{align}
\label{claseom}
\partial'_{t}\partial_{t}\phi_{a}(t,{\bf x})=\sum_{i}\partial_{i}'\partial_{i}\phi_{a}(t,{\bf x})+\mu^2\phi_{a}(t,{\bf x})-\frac{\lambda_0}{6\,N}\,
\left(\phi_{b}\phi_{b}\,\phi_{a}\right)(t,{\bf x}).
\end{align}
The classical simulations use the same lattice sizes and spacings
as in the quantum case. As observables we use
\be
\langle\phi(t,{\bf x})\phi(t,{\bf x})\rangle_{\rm class} \equiv
\left\langle\frac{1}{n_s^{3}N}\sum_{{\bf x}}\phi_{a}(t,{\bf x})\phi_{a}(t,{\bf x})
\right\rangle_{\rm class}
\ee
which is the analog of the quantum correlator
$F(t,\vecx;t,\vecx)=F(t,t,{\bf 0})$,
and the energy
\begin{align}
E=\sum_{{\bf x}}\left[\frac{\left(\partial_{0}\phi_{a}(x)\right)^2}{2}+
\frac{\sum_{i}\left(\partial_{i}\phi_{a}(x)\right)^2}{2}-
\frac{\mu^2}{2}\phi_{a}^{2}(x)
+\frac{\lambda_0}{24N}\left(\phi_{a}^2(x)\right)^{2} + V_0\right].
\end{align}
Comparing
$\langle \phi(t,{\bf x})\phi(t,{\bf x})\rangle_{\rm class}$
to
$F(t,\vecx;t,\vecx)$
we should keep in mind that the latter is divergent in the
continuum limit, whereas classically
\be
\langle\phi(0,{\bf x})\phi(0,{\bf x})\rangle_{\rm class}=
\frac{1}{n_s^{3}}
\sum_{|\veck|<\mu}
\frac{1}{2\sqrt{\mu^2 + k_{\rm lat}^2}} \simeq 0.0034.
\ee
We therefore compare
(cf.\ the discussion in the previous section)
\be
\label{Fphiphicomparison}
\langle \phi(t,{\bf x})\phi(t,{\bf x})\rangle_{\rm class}
\leftrightarrow F(t,\vecx;t,\vecx)- F(0,\vecx;0,\vecx).
\ee
ignoring the small difference
$\langle\phi(0,{\bf x})\phi(0,{\bf x})\rangle_{\rm class}\simeq 0.0034$
at time zero.

We also calculate occupation numbers and frequencies from the
$\phi\phi$ and $\pi\pi$ correlator in a manner similar to
eqs.\ (\ref{nkomdef},\ref{nkomdef2}) (without the `1/2'),
\be
\label{clasnk}
n_{k}=N^{-1}\sqrt{\langle\pi_{\bf k}^a\pi_{\bf -k}^a\rangle
\langle\phi_{\bf k}^b\phi_{\bf -k}^b\rangle},
\quad \omega_{k}=\sqrt{\langle\pi_{\bf k}^a\pi_{\bf -k}^a\rangle/
\langle\phi_{\bf k}^b\phi_{\bf -k}^b\rangle}.
\ee
We average over ${\bf k}$ with the same length $|{\bf k}|$. This
gives better statistics, and since the system is homogeneous and
isotropic as for the quantum case, the correlators
depend on the length of ${\bf k}$, apart from lattice artifacts.

In the classical approximation we expect
equipartition to occur after very long times,
for which the occupation numbers for the Goldstone and Higgs modes
are given by
\be
n_{k}^{\rm G}= \frac{T_{\rm class}}{k_{\rm lat}},
\quad
n_k^{\rm H} =\frac{T_{\rm class}}{\sqrt{m^2 + k_{\rm lat}^2}}.
\ee
The Rayleigh-Jeans temperature is determined by classical equipartition,
$E = N n_s^3 T_{\rm class}$, or with $E \approx V_0 L^3$,
$T_{\rm class} = (3/2) (\mu/\lm) (a_s\mu)^3$. In our longer-time
simulation we used $a_s\mu = 0.7$ and $\lm_0 = 6$, which gives the
low value $T_{\rm class} \simeq 0.086\, \mu$. The corresponding
guiding plots for the compound frequencies and particle numbers
are shown in figure \ref{omnguideclas}. The deviations from the BE
case in the compound-frequency plot appear limited but
significant. The greatest deviations appear in the
compound-particle number case: the curvature caused by plotting
$\ln(1+1/n_k)$ (instead of the more appropriate $1/n_k$), done
for diagnostic reasons, which cannot only be ascribed to a
dominating Goldstone contribution.

\FIGURE{
\epsfig{file=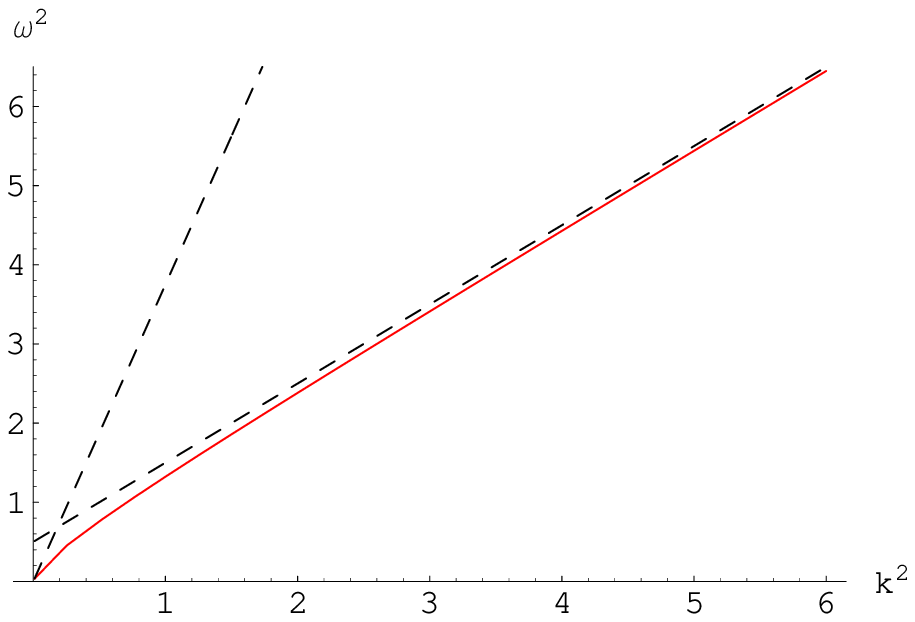,width=7cm,clip}
\epsfig{file=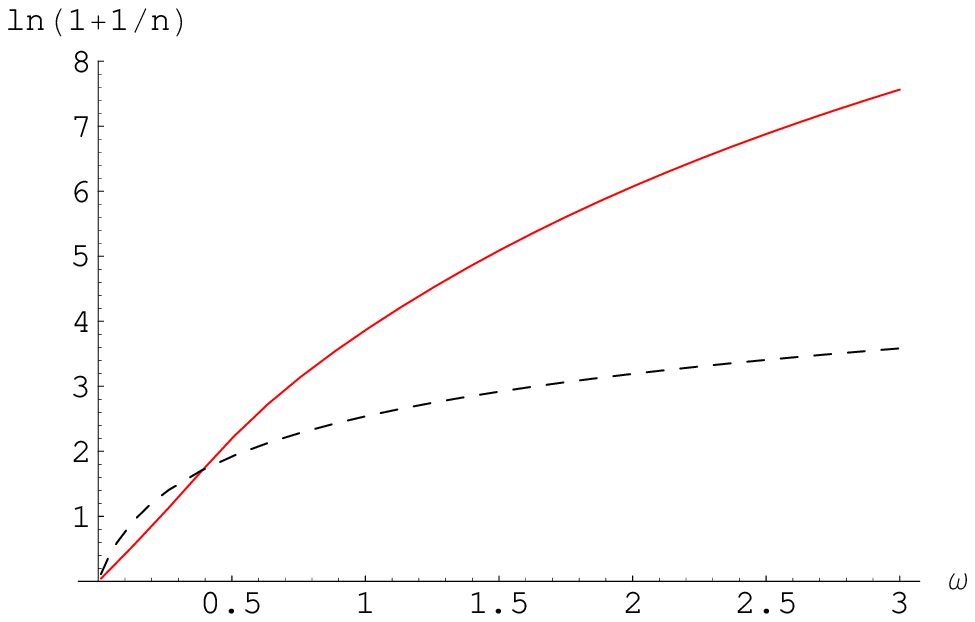,width=7cm,clip}
\caption{The Rayleigh-Jeans case,
$N=4$, $T=0.086\, \mu$, $m=\sqrt{2}\, \mu$, units $\mu=1$.
Left: Compound frequency spectrum, $\om_k^2$ versus $k^2$. The
dashed lines $\om^2 = 3.75\, k^2$ and $\om^2 = 0.5\mu^2 + k^2$ are
the asymptotes corresponding to (\ref{omas0}) and
(\ref{omasinfty}). Right: Compound particle number,
$\ln(1+1/n_k)$ versus $\om_k$.
The dashed line represents the pure Goldstone case,
$\ln(1+\om/T)$.
\label{omnguideclas}}
}

%
%

\section{Numerical results\label{results}}

\subsection{Early times: Tachyonic instability\label{sssecEarly}}

\FIGURE{
\epsfig{file=Early5.eps,width=10cm,clip}
\caption{The evolution in time of the equal time correlator eq.\
(\ref{Fphiphicomparison}) for NLO, Hartree, classical and LO. $\mu$ is used to set the scale.}
\label{Early5}
}
\FIGURE{
\epsfig{file=Early1.eps,width=14cm,clip}
\caption{
Compound occupation numbers $n_{k}$ vs $k/\mu$ for the Hartree
case (green/grey) during tachyonic preheating compared to the
quadratic approximation (black). The individual plots represent
different times.
Notice the logarithmic y-axis. The Hartree approximation
follows nicely the quadratic behaviour through the instability,
but back-reactions stop the growth of the modes. Very little power
is scattered into modes with $k>\mu$. }
\label{Early1}
}
We first describe simulations during the initial tachyonic
transition on a lattice with $n_s^{3}=32^{3}$ sites,
lattice spacing $a_s\mu=0.7$, volume $(L\mu)^3 = (22.4)^3$, time
step $dt=a_t/a_s=0.1$, with weak coupling $\lambda_0=1$. The
number of fields $N=4$.

We compare four different approximations: 1) The analytical result
in the quadratic approximation (eqs. (\ref{eqC1}), (\ref{eqC2}),
(\ref{eqC3})) ; 2) Hartree (cf. sect.\ \ref{Hartree}); 3) 1/N to
NLO (eqs.\ (\ref{eomFN}) -- (\ref{energy2PIN})); and 4) classical
(eq. (\ref{claseom})).
In the classical case, we average over an ensemble of 400
configurations. For the 1/N-NLO case relatively short times allow
us to keep the whole memory kernels in the self-energy.

In figure \ref{Early5} we show the evolution of the subtracted
$F(t,t,{\bf 0})$ for the NLO, Hartree
and classical (eq.\ (\ref{Fphiphicomparison})) case.
We have also included LO-1/N for comparison. The NLO and classical
agree well with each other (and settle later near the
zero-temperature vev,
$v^2/N = 6\,\mu^2$). The Hartree result is remarkably lower than the
others, while the classical approximation seems to work very well.
The Hartree being lower
appears to be the result of the choice of coefficient of the local
term, the choice of $N$. In the limit of large $N$ we recover the
LO result, which
for $N=4$ overshoots compared to the classical and NLO.
Below, we will discard the LO approximation, since it is
qualitatively the same as Hartree.

Figure \ref{Early1} shows the evolution of the
compound
occupation numbers
for the Hartree approximation (green/grey) compared to the
quadratic approximation (black). At early times the agreement is
very good but eventually back-reactions enter and the spinodal
growth
ends. As is well known, the homogeneous Hartree approximation does
not include
non-trivial scattering, and essentially no energy is
re-distributed to modes of momentum $k$ higher than $\mu$.
\FIGURE{
\epsfig{file=Early2.eps,width=14cm,clip}
\caption{As figure\ \ref{Early1} comparing NLO (black) to Hartree (green/grey).
At early times they agree well, but for NLO there is scattering
into higher momentum modes $k>\mu$.}
\label{Early2}
}

Figure\ \ref{Early2} compares Hartree (green/grey) to NLO (black) dynamics.
Initially they agree very well.
As the back-reaction kicks in, scattering processes included in the higher loop diagrams
of NLO
allow re-scattering of energy into
the higher momentum modes.  The spinodal growth ends roughly
 around $t\mu = 4-5$, when $F/\mu^2\simeq 4$ (cf.\ figure\
\ref{Early5}), at which time
$M^2(t)$ defined in (\ref{MdefN}) goes through zero (as anticipated in (\ref{stopest})).
\FIGURE{
\epsfig{file=Early3.eps,width=14cm,clip}
\caption{As figure\ \ref{Early1} comparing the classical approximation (green/grey)
to the NLO (black) result. Classically, only the unstable modes are initialized
and a continuous function is restored at times $\mu\, t\simeq 5$.
After the transition, the two approximations
agree
very well.}
\label{Early3}
}

In figure\ \ref{Early3} we compare the NLO (black) result to the
classical (green/grey) case. In the classical case only the
unstable modes are initialized, which is why the stable modes are
outside the picture initially (having $n_{k}=0$). In the quadratic
approximation the evolution for classical modes and quantum mode
functions is identical, so we expect the unstable modes to also be
described by eq. (\ref{spinnk}). This is indeed what we see, and
as back-reactions kick in, the stable modes become populated and
``hook up'' with the unstable ones to reproduce the low-momentum
part of the distribution of the NLO case very well.
Turning things around,
the {\it full} dynamics of the classical evolution is well described by the
{\it truncated} NLO quantum dynamics.
\FIGURE{
\epsfig{file=Early4.eps,width=14cm,clip}
\caption{The dispersion relation ($\omega_{k}^{2}/\mu^{2}$ vs $k^{2}/\mu^{2}$) in the NLO (green/grey) case compared to the quadratic approximation (black).}
\label{Early4}
}

Figure\ \ref{Early4} shows the
compound-dispersion relation during the transition at NLO. The low
momentum modes are again well approximated by the gaussian
equations (\ref{eqC1}, \ref{eqC2}, \ref{eqC3}), and although the
quantitative agreement is lost for large momentum modes and larger
times, the qualitative picture is well understood. For the Hartree
case the picture is roughly the same at these early times. The
averaging in the classical case was seen to smear the oscillations.

%
%

\subsection{Intermediate times:
approach to equilibrium\label{secthermalization}}
\FIGURE{
\epsfig{file=Late2.eps,width=14cm,clip}
\caption{The evolution of the dispersion relation ($\omega_{k}^{2}/\mu^2$ vs $k^{2}/\mu^2$) in time for NLO, classical and Hartree (top to bottom). Overlaid a straight line with slope 1 and intercept 0 (free massless propagator).}
\label{Late2}
}
%


%
\FIGURE{
\epsfig{file=Late1.eps,width=14cm,clip}
\caption{The evolution of $\ln{\left(1+1/n_{k}\right)}$ vs. $\omega_{k}/\mu$ in time. From top to bottom NLO, classical and Hartree.
}
\label{Late1}
}
\FIGURE{
\epsfig{file=Late3.eps,width=7cm,clip}
\epsfig{file=Late4.eps,width=7cm,clip}
\caption{At the latest time $\mu\, t=700$, the
NLO compound dispersion relation and particle numbers
begin to show the expected behaviour, as in figure \ref{omnguide}. The overlaid lines are (left) asymptotes as in figure \ref{omnguide}  and (right) a straight-line fit as described in the main text.
}
\label{Late4}
}
As interactions become important the spinodal transition ends. We may expect scattering to result in equilibration of the system.
This does not happen in the Hartree approximation, but
the inclusion of the NLO diagrams
has been seen in previous studies to lead to thermalization,
in 1+1, 2+1 with scalar fields and in 3+1 dimensions including fermions \cite{AaBo00,Juchem:2003bi,Berges:2002wr,Berges:2004ce}.
In the present case, however, the suppression of the scattering of
the massless Goldstone bosons at low momenta (cf.\ (\ref{GBscat})
may lead to slowing down of the thermalization of these modes.
In the classical approximation, we expect
classical equipartition to take place after a very long time. For
intermediate times, it was seen in \cite{Skullerud:2003ki} that
the low momentum modes of a system including $SU(2)$ gauge fields
could be well approximated by a Bose-Einstein distribution,
allowing for a determination of an effective temperature and
chemical potential, even
with classical dynamics.

To study the longer time behavior we performed
simulations on a smaller lattice, $n_s^{3}=16^{3}$,
again with $\mu=0.7$, $dt=0.1$, $N=4$, but with a
larger coupling $\lambda_0=6$.
For the classical simulations we averaged over 1000 realizations
of the initial conditions.
For the quantum NLO case we cut off the memory kernel at $38.5\, \mu^{-1}$.
This gave
no problems with energy conservation and we believe
it
does not affect the evolution much. With the enhanced back-reaction of the larger coupling we expect smaller $n_{k}$ after the transition, since
the spinodal transition does not last as long.
Instead of a zero mode reaching previously $n_{0}\sim 1000$, we have here only
$n_{0}\sim 200$. This still amply satisfies the criterion $n_k\gg1$ for
classicality.

In figure\ \ref{Late2} we see the evolution of the compound dispersion relation.
In the NLO case, the oscillating behavior generated during the transition (figure \ref{Early4}) is washed out by scattering to a straight line.
A close-up at the latest time the NLO compound-frequency plot
figure\
\ref{Late4} appears to show the behavior anticipated in figure
\ref{omnguide}. The asymptote to the data at larger momenta
indicates an effective Higgs mass  $m^{2}\approx 1.80\, \mu^{2}$
(cf.\ (\ref{omasinfty})), somewhat lower than the zero-temperature value. At finite temperature in the broken phase we do expect a smaller Higgs mass, since the condensate is also smaller than at zero temperature (see figure \ref{Cond}, left). Inserting the finite temperature $v^{2}$ in the expression for the mass $m^{2}=\frac{\lambda}{3N}v^{2}$, one finds $m^{2}\approx 1.76\, \mu^{2}$.
\FIGURE{
\epsfig{file=Cond.eps,width=7cm,clip}
\epsfig{file=Tvt.eps,width=7cm,clip}
\caption{Left:The evolution of the quantity $v^{2}/\mu^{2}$ as defined in (\ref{condest}). Overlaid the zero temperature value (\ref{vest}). Right: The effective temperature in the low momentum range (from linear fits as in figure \ref{Late4}, right). Overlaid, an exponential fit.
}
\label{Cond}
}

The slope of the line through the origin and the first data point
is at this time $1.22$, quite a bit off the anticipated  $1.395$ (see figure \ref{omnguide}). With this slope and effective mass, no choice of $T$ satisfies (\ref{omas0}). Apparently we are not sufficiently close to equilibrium, in particular in the low-momentum region. 
Figure \ref{Late1} shows the evolution of the
compound
occupation numbers at later times. At NLO,
redistribution is clearly taking place, and power is being moved
towards higher momentum modes. In the Hartree approximation
nothing much happens.
The classical case also shows redistribution over the modes and
mimics the quantum evolution nicely at low momenta for not too
long times.
At the latest time some concaveness appears to show
up, perhaps indicating classical equipartition (cf.\ figure\
\ref{omnguideclas}).
We come back to this in more detail in the next section.
In figure \ref{Late4} we show a close-up of the NLO case at the latest
time. We include a fit to the low momentum region (neglecting the zero mode). As discussed in section \ref{ssecobservables} (see figure \ref{omnguide}), such a straight line represents a thermalized state, with the slope equal to $1/T$, and intercept equal to the effective (Goldstone) chemical potential. For this fit we find $T\simeq 1.1\, \mu$ and $\mu_{\rm ch}^{\rm G}\simeq 0.32\, \mu$. The temperature is higher than our free-field estimate (\ref{Test}). This is not surprising since the high-momentum range is still under-populated at this time, compared to the thermal state. Figure \ref{Cond}, right, shows the evolution of this temperature in time. An exponential fit suggests an asymptotic final temperature of around $T/\mu=1.04$.

%
%

\subsection{Choice of classical initial conditions\label{whichhalf}}

\FIGURE{
\epsfig{file=Choice1.eps,width=9cm,clip}
\caption{Comparing the NLO result to the classical with unstable modes
and all modes initialized. $\mu\, t=14$, just after the tachyonic transition
($L\mu = 22.4$, $\lm_0=1$).}
\label{Choice1}
}

\FIGURE{
\epsfig{file=Qvcl.eps,width=7.4cm,clip}
\epsfig{file=Choice2.eps,width=7.4cm,clip}
\caption{Comparing the NLO result to the classical with unstable modes
and all modes initialized.
Left:
$\mu\, t=100$, after some equilibration. Right:
$\mu\, t=700$.
($L\mu=11.2$, $\lm_0=6$). }
\label{Choice2}
}
As an aside we comment on the choice of initial conditions for classical simulations of tachyonic preheating. Two approaches have been put forward,
one used by many authors (see e.g.\ \cite{Rajantie:2000nj,Copeland:2002ku}),
in which
{\em all} modes are initialized with the ``quantum half'',
i.e.\ also $k>\mu$ in eq.\ (\ref{quanthalf}),
the other
\cite{Salle:2001xv,Smit:2002yg,Tranberg:2003gi,Garcia-Bellido:2002aj,Garcia-Bellido:2003wd},
in which only the unstable modes are initialized,
as we did in the previous section.

It is clear that initializing all the modes, we risk running into an ultraviolet problem in the
continuum limit $a_{s}\rightarrow 0$.
In the classical approximation,
equipartition may cause the divergent energy sitting in the high
momentum modes to flow into the low-momentum range and generate a
too high temperature, see \cite{Salle:2001xv}, and
\cite{Moore:2001zf} for a study with massless gauge fields.
However, this did not happen on short to intermediate time scales
in our $SU(2)$-Higgs study in which the gauge fields were massive
\cite{Skullerud:2003ki}, and also for scalar fields with only
self interactions the problem shows up on a much larger time scale
than for massless gauge fields \cite{Moore:2001zf}.
Thermalization for scalar fields is rather slow
and it is possible that the
high-momentum
modes never come into play on time-scales relevant to numerical simulations of
phase transitions.
The important criterion must be which initialization method reproduces the quantum evolution of the low momentum modes more closely, in terms of occupation numbers and equilibration times. We shall take the NLO case as representing the quantum result.

It is also
important to keep in mind that, when attempting to calculate
topological quantities such as
the Chern-Simons number in models which include gauge fields, the lattice implementation
is very sensitive to ultraviolet fluctuations. In simulations of baryogenesis, Chern-Simons number is a crucial observable, since it is simply related to the generated baryon asymmetry. This was in fact one of the motivations for introducing the initialization restricted to the unstable modes
\cite{Smit:2002yg,Tranberg:2003gi}.

When initializing all modes, the effective mass receives a
contribution from all these modes (divergent in the continuum limit),
compared to when only the unstable modes are initialized,
in a way analogous to the quantum vacuum case. We should therefore
introduce a counterterm for the mass in a way similar to the
quantum case \cite{Rajantie:2000nj},
\be
\label{clasren}
-\mu^2\rightarrow -\mu^2-\lambda_0\frac{N+2}{6N}c_{1}(\mu),
\ee
where again $c_{1}$ is given by eq.\ (\ref{c1def}). For consistency, we
also define the occupation numbers according to the quantum equation,
\be
n_{k}+1/2=N^{-1}\sqrt{\langle\pi_{\bf k}^a\pi_{\bf -k}^a\rangle
\langle\phi_{\bf k}^b\phi_{\bf -k}^b\rangle},
\ee
rather than through eq.\ (\ref{clasnk}),
and the energy and pressure also have to be renormalized.

The result is shown in figure\ \ref{Choice1}, where the two
classical approaches are compared to the NLO quantum case. Just
after the transition ($32^{3}$, $\lambda_0=1$), the
classical with only the unstable modes initialized follows the
quantum case nicely in the low momentum region. This is also the
case when initializing all the modes. We conclude that as long as
the simple renormalization procedure eq.\ (\ref{clasren}) is
imposed, the number of modes initialized is not crucial at these
early times,
when
classical equilibration has not set in yet.

For longer time (figure\ \ref{Choice2} left plot, $16^{3}$, $\lambda_0=6$, $\mu\,t=100$),
we see that both types of initialization approximate the quantum
behavior, with the `all initialized' doing somewhat better
than the `unstable only'. 
We do not expect the classical modes with occupation numbers 
less than one to be correct.
Surprisingly, the very lowest modes with $n_k \gg 1$ are also off in both cases.
At 
even longer times (figure\ \ref{Choice2} right plot, $\mu\,t=700$), the agreement is only qualitative,
 with either intitialization scheme performing badly.
For only the unstable modes initialized, we see that the
occupation numbers are lower than for NLO, presumably because of
the lack of scattering with the high momentum (stable) modes.
In the quantum case there is apparently still some scattering
going on with the
quantum fluctuations of the initial free `vacuum'. When
initializing all the modes,
this effect appears to be incorporated in the `all initialized',
but it seems to be too large. This could also be the reason that
the zero mode is lower than for NLO. Eventually, we expect $n_0$
to approach the equilibrium value following from
(\ref{Fguide}, \ref{Fddguide}), $n_0\approx \left(v^2 L^3 (n_0^{\rm
H} +1/2) m/N^2\right)^{1/2}-1/2 \simeq 19$, for the parameters of
the simulation.

\FIGURE{
\epsfig{file=Choice3.eps,width=10cm,clip}
\caption{The occupation numbers in the high momentum modes for NLO, classical with unstable initialized and classical, all initialized. $\mu \,t=700$. For the classical, all, $n_{k}<0$ for the highest modes, which means that $n_{k}+1/2<1/2$.} \label{Choice3}
}

There is an interesting aspect to this. In figure\ \ref{Choice3} we see
the high momentum modes at the same time as figure\
\ref{Choice2}, right; shown is $n_{k}$ vs $k/\mu$ on a linear scale.
In the quantum case, the quantity $(n_{k}+1/2) -1/2$ is always
positive and here close to zero, as we would expect. This is a
consequence of the standard quantum commutation relation. The
power in the vacuum fluctuations, the ``$1/2$'', cannot be
extracted and moved to other modes.  For the classical case with
only unstable modes initialized, the $n_{k}$ are
also manifestly positive numbers, and close to zero. In the
classical
case with all modes initialized, there is no principle which
enforces that $(n_{k}+1/2)$ should be larger than $1/2$, and we
see that it ends up negative for $k \gtrsim 3.5\, \mu$. This illustrates
the problems associated with renormalization in classical field
theory out of equilibrium. Only for short times, as we have seen,
or {\em in} equilibrium \cite{Tang:1996qx,Tang:1998gk,AaSm97},
such a procedure has a limited validity, albeit with initial-state
or temperature-dependent counterterms.

\subsection{Effective temperature in the unstable modes}

In \cite{Garcia-Bellido:1999sv,Garcia-Bellido:2003wd} 
the baryon asymmetry in scenarios
based on electroweak preheating was estimated using 
equilibrium concepts, such as the sphaleron rate at an effective 
temperature and an effective chemical potential for Chern-Simons number. 
The idea was that the large 
occupation numbers in the unstable modes would lead to early equilibration 
amongst these modes at an effective temperature $T_{\rm eff}$. 
Such an approach might enable one to bypass
the more difficult numerical simulations using CP violation in the equations
of motion (we note that the comparison made in \cite{Tranberg:2003gi} 
did not seem to support this idea at a quantitative level). 
It is therefore of interest to estimate such an effective temperature here.

Equilibrated modes with high occupation numbers have an
effective temperature given by the classical formula 
$T_{\rm eff} = n_k \om_k$, independent of $k$, and one estimate of the
effective temperature follows from energy conservation and dominance
of the unstable modes \cite{Garcia-Bellido:1999sv}, 
\be
N\int_{k < \mu}\frac{d^3 k}{(2\pi)^3}\, n_k \om_k
= \frac{N\mu^3 T_{\rm eff}}{6\pi^2} = V_0 
= \frac{3N\mu^4}{2\lm} 
\quad\Rightarrow 
T_{\rm eff} = \frac{9\pi^2}{\lm}\,\mu
= \frac{9\pi^2}{\sqrt{6 N \lm}}\, v.
\label{TeffGBetal}
\ee
With $v=246$ GeV, $N=4$, this gives a high 
$T_{\rm eff}\simeq 4460$ GeV (1820 GeV), for $\lm=1$ (6). 
The actual $n_k \om_k$ of the individual modes as obtained from the
$\langle\partial_t\ph \partial_t\ph\rangle$ correlator ($1/N$ Higgs- 
plus $(N-1)/N$
Goldstone-contributions, as in (\ref{Fddguide})), 
is shown in figure \ref{figTeffmodes} (left)
at time $t\mu = 14$, for the $32^3$ lattice with $\lm=1$.
We see that the
unstable modes have not equilibrated much yet, since the effective
temperatures vary significantly from mode to mode
(the zero mode should be ignored as it contains the condensate). Note that
the corresponding particle distribution is shown in the lower-right panel 
of figure \ref{Early3}. As the panel $t\mu = 2.8$ in the latter 
figure shows, the modes near the border of the unstable region
($k\approx \mu$) are not initially highly populated, and also at
time $t\mu=14$ their effective temperature is much lower than the simple
estimate (\ref{TeffGBetal}). 
The effective temperature in the modes near $k/\mu = 0.5$ 
in figure\ \ref{figTeffmodes} (left) is about 100 $\mu$; the
average over all unstable (non-zero) modes gives 
$\overline{T_{\rm eff}} = \sum_{\veck, k<\mu} n_k\om_k/
\sum_{\veck, k<\mu} = 54\mu$, about half the estimate 
(\ref{TeffGBetal}) for $N=4$, $\lm=1$.
\FIGURE{
\epsfig{file=nkwk_l1_t14.eps,width=7cm,clip}
\epsfig{file=Teff_16_in_t.eps,width=7cm,clip}
\caption{Left: effective temperatures 
$T_{{\rm eff}, k}\equiv n_k\om_k$ (in units of $\mu$)
of the unstable modes of the $32^2$ lattice 
at time $t\mu=14$; the coupling $\lm=1$.
Right: same for the $16^3$ lattice (still with $\lm=1$), 
as a function of time. 
Top to bottom: zero mode, $L |\veck|/(2\pi)=1$, $\sqrt{2}$, $\sqrt{3}$.
}
\label{figTeffmodes}
}
The effective temperatures at later times are shown in figure
\ref{figTeffmodes} (right). Equilibration is evidently slow even 
in the unstable
momentum region, which is perhaps to be expected because of the
suppressed scattering of Goldstone bosons at low momenta. 
In a more realistic setting, the gauge fields
and other degrees of freedom of the SM will lead to more efficient
thermalization and lower effective temperatures. 
In the study \cite{Skullerud:2003ki} the SU(2) gauge 
fields were
taken into account, which led to effective temperatures 
$T_{\rm eff} \approx 0.25 - 0.4\, m_H\approx 0.4-0.6\, \mu$ at $t m_H \approx 30 - 40$ 
(cf.\ figure.\ 15 and Table 2 in
ref.\ \cite{Skullerud:2003ki}), for $m_H = \sqrt{2}\, m_W$.

%
%

\section{Conclusion}

The $\Phi$-derived $1/N$--NLO approximation performed well in this
study of a tachyonic phase transition, for the $O(4)$ model at
reasonable couplings\footnote{At stronger couplings 
$\lm \to 10$ we did see
annoying instabilities similar to what we found in approximations
based on the weak-coupling expansion of $\Phi$. In this sense the
larger coupling $\lm=6$ used in Sect.\ \ref{results}
is not very weak.}. It is reassuring
that the classical approximation agrees so well with the quantum
case, not only during the transition, were we used a smaller
coupling (with the accompanying larger particle numbers), but also
subsequently for not too long times, where we used the larger
coupling. In a sense, the classical and NLO approximations support
each other, the first taking into account all non-linearities
without further truncation and the second including quantum
scattering right from the beginning. For longer times the quantum
computation is of course superior as it does not suffer from
Rayleigh-Jeans ambiguities.\footnote{For some problems one may
have to face a non-uniformity of large
$N$ versus large $t$ \cite{Ryzhov:2000fy}.}

Because of the $O(N)$ symmetry of the initial state, the mean
field was zero at all times. Consequently there was no
conflict with Goldstone's theorem -- we consider the numerical
evidence for massless particles to be quite convincing. 

Critical slowing-down of thermalization is expected to take place
as a result of the suppressed interactions of Goldstone bosons at low
momenta. We do see at late times that the intermediate range of momenta approach a thermal form, while the very lowest ones may be suffering from this critical slowing down. A larger volume (providing very low momentum modes) is needed to firmly establish this. In finite volume one would anyway expect a non-zero mean field to vanish as the system approaches equilibrium. In this explorative study we did not pursue the finite-volume aspects in great detail and more can be done in a finite-size scaling analysis. 

\subsection*{Acknowledgments}
We thank Gert Aarts, J\"urgen Berges, Szabolcs Bors\'anyi,
and Julien Serreau for useful discussions, and the referee of this paper for
useful suggestions. This work was supported by FOM/NWO, 
and was supported in part by the NSF Grant No.\ PHY94-07194.

\bibliographystyle{JHEP}
\bibliography{anderslit2JS}
\end{document}

%% file: newcomm.tex
\newcommand{\intx}{\int d^4 x\,}

\newcommand{\veck}{{\bf k}}

\newcommand{\vecx}{{\bf x}}       

\newcommand{\vecz}{{\bf z}}

\newcommand{\dl}{\delta}
\newcommand{\ep}{\epsilon}

\newcommand{\lm}{\lambda}
\newcommand{\rh}{\rho}
\newcommand{\sg}{\sigma}

\newcommand{\ph}{\phi}

\newcommand{\om}{\omega}

\newcommand{\Gm}{\Gamma}

\newcommand{\Sg}{\Sigma}

\newcommand{\half}{\frac{1}{2}}

\newcommand{\Tr}{\mbox{Tr}\,}

\newcommand{\eela}[1]{\label{#1}\end{equation}}
\newcommand{\eeala}[1]{\label{#1}\end{eqnarray}}
\newcommand{\bea}{\begin{eqnarray}}
\newcommand{\eea}{\end{eqnarray}}